# Theoretical study of the role of pairing symmetries on the physical parameters of Fe-based superconductors

Sushree Sangita Jena[1], G. C. Rout[2*] and S. K. Agarwalla[3]

[1,3] Department of Applied physics and Ballistics, F.M.University, Balasore 756019, Odisha, India,

[2] Condensed Matter Group, Physics Enclave, Flat No-2H, Block-2, Krishna Garden Annex Apartment, Jagamara, Po-Khandagiri, Bhubaneswar-751030, Odisha, India.

[*] Corresponding author: Email Id: gcr@iopb.res.in

**Abstract.** For the newly discovered iron-based superconductors, we propose a one band model for conduction band with first and second-nearest-neighbor electron hoppings with d-wave, $s_{x^2+y^2}$-wave, $s_{x^2y^2}$-wave, $s_0$-wave and $d_{x^2-y^2}$-wave pairing symmetries for the superconductivity. We have calculated the Green's functions from which the temperature dependent superconducting (SC) gap and tunneling conductance are calculated and computed numerically taking 100×100 grid points of electron momentum throughout the Brillouin zone. The evolution of the SC gap and the tunneling conductance spectra are investigated for different model parameters of the system.

## 1. Introduction

The superconductivity was discovered for the first time in the iron-based oxypnictide LaFeAsO (F doped) of the type-1111 with critical transition temperature $T_c$ = 26 K [1]. Then the superconductivity was discovered in a series of 122-type materials having Fe-As layers: such as $AFe_2As_2$ (with A= K, Na, Rb) and $MFe_2As_2$ (with M = Ca, Ba, Sr) [2, 3, 4]. At present, the highest reported $T_c$ = 50 K for iron-based superconductors was achieved in SMFeAsO and the superconductivity in the Fe-based materials is unconventional and non-BCS type [5, 6, 7]. Guo et. al. have communicated the superconductivity at $T_c$ = 30 K in FeSe-layer for the compound $K_xFe_2Se_2$ [8]. The angle resolved photoemission spectroscopy (ARPES) experiments [9, 10, 11, 12, 13, 14, 15] and Local density approximation (LDA) calculations [16, 17, 18] have communicated that the electron Fermi surface (FS) pocket only exists, but the hole Fermi surface disappears. This compound may be heavily electron doped superconducting material. The pairing mechanism in these systems is still unclear for the systems and d$_{x^2-y^2}$-wave, s$_\pm$ = s$_{x^2y^2}$-wave and s-wave pairing symmetries have been proposed [19, 20, 21, 22, 23, 24, 25, 26, 27]. The band structure calculations [16, 17, 18] of the compounds show that all the five 3d orbitals of iron atoms hybridize strongly to contribute to the electronic density of states. To simplify the problem, Raghu. et. al. [28, 29] have proposed a two band model.

We propose here a one band tight binding model for the conduction band with the nearest-neighbor (NN) and the next- nearest-neighbor (NNN) electron hoppings and superconducting interaction with different pairing symmetries for these systems. We attempt here to study the temperature dependent superconducting gap and conductance spectra measured by several experiments. The theoretical model incorporating NN and NNN- hopping interactions and the k-dependent superconducting gap interactions is presented in section 2. The calculation of superconducting gap by Green's function technique is given in section 3. We present the calculation of the electron specific heat in section 4 and the electron density of states in section 5. We report the results and discussion of the 122-systems in section 6 and finally the conclusions is given in section 7.

## 2. Formalism

It is reasonable to believe that the electronic bands crossing the Fermi surface (FS) are very important to construct a minimal tight binding model, while other bands not crossing the Fermi energy can be ignored. The unit cell of the system consists of two iron atoms. Due to iron vacancy, only one kind of Fermi surface and its symmetry near the Fermi surface in the Brillouin zone is essential. Therefore, we consider here the one band model taking the nearest-neighbor (NN) and next-nearest-neighbor (NNN) electron hoppings. The model Hamiltonian also consists of an electron momentum (k) dependent superconducting gap interaction with different possible pairing symmetries. The model Hamiltonian for the 122-type family is described as,

$$H = \sum_{k,\sigma} \varepsilon_k C_{k\sigma}^+ C_{k\sigma} + \sum_k \Delta(k) (C_{k\uparrow}^+ C_{-k\downarrow}^+ + C_{-k\downarrow} C_{k\uparrow}) \qquad (1)$$

Where $\varepsilon_k$ is the single band dispersion with the NN and the NNN-hopping interactions in the Fe-Se plane of the system. The band energy for this single band model of the iron based superconducting system is given by,

$$\varepsilon_k = -2t_1(\cos kx + \cos ky) - 42t_2 \cos kx \cos ky \qquad (2)$$

Here the electron hopping integrals correspond to the NN- and NNN- hopping interactions and the components of the electron momentum $\vec{k}$ are the kx and ky. The momentum dependent superconducting (SC) gap, $\Delta(k)$ in the Fe-Se plane is given by,

$$\Delta(k) = \sum_{\acute{k}} V(k - \acute{k})(C_{\acute{k},\uparrow}^+ C_{-\acute{k},\downarrow}^+) \qquad (3)$$

where $V(k - \acute{k})$ is the momentum dependent effective Coulomb potential responsible for the formation of the Cooper pairs. There are three kinds of pairing symmetries namely the d-wave pairing symmetry with SC gap $\Delta(k) = \Delta_0(T)(\cos kx - \cos ky)$ the $s_\pm$ wave pairing symmetry with $\Delta(k) = \Delta_0(T)(\cos kx \cos ky)$ and the isotropic s-wave symmetry with $\Delta(k) = \Delta_0(T)$.

## 3. Calculation of the superconducting gap equation

The Hamiltonian written in equation (1) for the iron-based superconductors is solved by calculating the electron Green's functions. For this we evaluate the equations of motion of the fermion operators. The fermion operators involved are found to be,

$$i\frac{dC_{k\uparrow}}{dt} = [C_{k\uparrow}, H] = \varepsilon_k C_{k\uparrow} + \Delta(k) C^+_{-k\downarrow} \tag{4}$$

$$i\frac{dC^+_{-k\downarrow}}{dt} = [C^+_{-k\downarrow}, H] = -\varepsilon_k C^+_{-k\downarrow} + \Delta(k) C_{k\uparrow} \tag{5}$$

For the solution of the Hamiltonian, two electron Green's function are involved in the calculation and their Fourier transformed Green's functions are given as,

$$A_1(k,\omega) = \langle\langle C_{k\uparrow}; C^+_{k\uparrow} \rangle\rangle_\omega \tag{6}$$

$$A_2(k,\omega) = \langle\langle C^+_{-k\downarrow}; C^+_{k\uparrow} \rangle\rangle_\omega \tag{7}$$

These Green's functions are calculated for the Hamiltonian using Zubarev's Green's functions technique [30] and the coupled equations are written as,

$$(\omega - \varepsilon_k) A_1 = \frac{1}{2\pi} + \Delta(k) A_2 \tag{8}$$

$$(\omega + \varepsilon_k) A_2 = \Delta(k) A_1 \tag{9}$$

Solving these two coupled equations given in equations above, we finally obtained these two Green's functions,

$$A_1(k,\omega) = \frac{1}{2\pi} \frac{(\omega + \varepsilon_k)}{\omega^2 - E_k^2} \tag{10}$$

$$A_2(k,\omega) = \frac{1}{2\pi} \frac{\Delta(k)}{\omega^2 - E_k^2} \tag{11}$$

Where $E_k = \pm\sqrt{\varepsilon_k^2 + \Delta^2(k)}$ where the electron dispersion $\varepsilon_k$ involving the two hopping parameters $t_1$ and $t_2$ is defined earlier in equation (2) and the electron momentum dependent SC gap parameter $\Delta(k)$ is also defined earlier in equation (3) for describing different pairing symmetries. Before calculating the expression for the SC gap parameter, we evaluate the electron correlation function associated with the

Green's function $A_2(k,\omega)$ written in equation (5). The correlation function is defined as,

$$\langle C^+_{k\uparrow} C^+_{-k\downarrow}\rangle = \int_{-\infty}^{\infty} J_k(\omega)\, e^{-i\omega(t-\acute{t})} d\omega \quad for\ t = \acute{t} \tag{12}$$

where the spectral intensity $J_k(\omega)$ is written as,

$$J_k(\omega) = i \lim_{\eta\to 0} f(\beta\omega)[A_2(\omega + i\eta) - A_2(\omega - i\eta)] \tag{13}$$

where the Fermi-Dirac distribution function is given by, $f(\beta\omega) = [1 + \exp(\beta\omega)]^{-1}$ with $\beta = 1/k_B T$. After simplification, the correlation function appears as,

$$\langle C^+_{k\uparrow} C^+_{-\acute{k}\downarrow}\rangle = -\frac{\Delta(\acute{k})}{2E_{\acute{k}}} \tanh\left(\frac{1}{2}\beta E_{\acute{k}}\right) \tag{14}$$

The momentum dependent SC gap $\Delta(k)$ defined earlier in equation (3) is rewritten as,

$$\Delta(k) = \sum_{\acute{k}} [V(k - \acute{k})\langle C^+_{\acute{k}\uparrow} C^+_{-\acute{k}\downarrow}\rangle] \tag{15}$$

where the momentum dependent effective Coulomb interaction is defined as,

$$V(k - \acute{k}) = V_q = -V_0 f(\acute{k}) f(k) \tag{16}$$

where $V_0$ here taken as the effective attractive Coulomb interaction between the electrons and $f(k)$ and $f(\acute{k})$ represent the different pairing symmetries associated with different types of superconductors. Further, the momentum dependent SC gap is defined as, $\Delta(\acute{k}) = \Delta_0(T) f(k)$, where $\Delta_0(T)$ represents the temperature dependence SC gap parameter. Employing these conditions and using the correlation function given in equation (14), finally the SC gap equation appears as,

$$1 = V_0 \sum_k \left[\frac{f^2(k)}{2E_k} \tanh\left(\frac{1}{2}\beta E_k\right)\right] \tag{17}$$

The momentum summation appearing in gap equation (17) in 2D electron momentum plane is written as, $\sum_k \frac{S}{(2\pi)^2} \iint_{-\pi}^{\pi} dk_x dk_y$. Further the SC coupling constant is defined as, $g = V_0/t_1$, where $t_1$ is the nearest-neighbor electron hopping. The temperature dependent gap equation (17) is solved numerically taking $100 \times 100$ grid points in the xy-plane of the electron momentum. The results are discussed in the next section.

## 4. Calculation of Electron Specific heat

In order to calculate the electron specific heat and the entropy of iron-based superconductor, we set up below the Fermi energy of the system in terms of the quasi-particle band energy, $\omega_{\alpha k}$ with $\alpha = 1$ and 2 for the two bands. The free energy is written as,

$$F = -k_B T \sum_{\alpha,k,\sigma} [\ln\{1 + \exp(-\beta\omega_{\alpha k})\}] \tag{18}$$

where $\beta = 1/k_B T$. The electronic entropy of the system in the superconductivity state can be calculated using the given formula,

$$S = -\frac{1}{N_A}\left(\frac{\partial F}{\partial T}\right)_{V,\mu} \tag{19}$$

where A, V, μ are respectively is the number of number of atoms, the volume and the chemical potential.

After evaluation from the free energy the entropy becomes,

$$\frac{S}{k_B} = \frac{1}{N_A} \sum_{\alpha,k,\sigma} [\ln\{1 + \exp(-\beta\omega_{\alpha k})\} + \frac{1}{\{\exp(-\beta\omega_{\alpha k})+1\}} \{\beta\omega_{\alpha k} - \omega'_{\alpha k}\}] \qquad (20)$$

Here $\omega'_{\alpha k}$ written as, $\omega'_{\alpha k} = \frac{\partial \omega_{\alpha k}}{\partial (k_B T)}$ for $\alpha = 1$ and 2. The explicit evaluation of the differential of the quasi-particle band reduces to,

$$\omega'_{1k}, \omega'_{2k} = \frac{\Delta(k)\, \dot\Delta(k)}{\sqrt{\varepsilon_k^2 + \Delta^2(k)}} \qquad (21)$$

The momentum and temperature dependent superconducting gap $\Delta(k)$ is differentiated with respect to temperature. So that we obtain $\dot\Delta(k)$ which is defined as,

$$\dot\Delta(k) = \frac{\partial \Delta(k)}{\partial (k_B T)} \qquad (22)$$

We have numerically and self-consistently evaluated these two quantities $\Delta(k)$ and $\dot\Delta(k)$ before numerical calculation of entropy and the electron specific heat. The electron specific heat is evaluated from entropy S, is given below at constant volume V and chemical potential μ,

$$\frac{C_V}{k_B} = T \left[\frac{\partial S}{\partial (k_B T)}\right]_{V,\mu} \qquad (23)$$

## 5. Caculation of electron DOS

The electron DOS describes the tunneling spectra of the iron-based superconductor. The electronic DOS is defined as,

$$DOS = -2\pi \sum_{k,\sigma} [\,Im\, A\,(k,\omega + i\eta)] \qquad (24)$$

Here $\eta$ is the small spectral width assigned to the frequency $\omega$ and $A\,(k,\omega)$ is the electron Green's function given in equation (5.15). After evaluating the imaginary part of $A\,(k,\omega)$ from equation (5.15), the DOS reduces to,

$$DOS = 2 \times \frac{S}{(2\pi)^2} \times \int_0^{2\pi}\int_0^{2\pi} dk_x\, dk_y \left\{\frac{(\omega + \varepsilon_k)2\eta\omega}{(\omega^2 - E_k^2 - \eta^2) + 4\eta^2\omega^2}\right\} \qquad (25)$$

Here we have converted the summation over 'k' into integral relation in the momentum x and y-plane of the electron momentum. Here '2' appear for the two spin orientation of the orbital and 'S' is the area of the unit cell in the real space.

The electron momentum spans over the Brillouin zone in the two-dimensional square lattice. The summation for electron momentum appearing in equation (24) is changed to the integral form in the momentum plane and appears as $\sum_k \to \frac{S}{(2\pi)^2} \iint dk_x dk_y$, where 'S' is the area of of the square lattice. The integration is carried out for 100×100 grid points of the kx and ky components of the electron momentum. The dimensionless parameters (scaled by nearest neighbour-hopping parameter $t_1$= 0.02 eV) are written as: NNN electron hopping integral, $t_2 = -0.045$, superconducting (SC) gap, (with temperature dependent gap $\Delta_0(T)$), temperature $t = k_B T/t_1$ and the SC coupling $g = V_0/t_1$ (with $V_0$ as the momentum independent effective Coulomb energy) and band energy $c = \omega/t_1$.

## 6. Results and Discussion

### 6.1 Temperature dependent study of SC gap

We have taken the NN and the NNN electron hopping parameters in conduction band dispersion, momentum dependent Coulomb potential $V(k - \acute{k})$ and different pairing symmetries for the superconducting (SC) gap and computed the temperature dependent superconducting gap parameter, $\Delta_0(T)$ given in equation (17) numerically and self-consistently to achieve the same SC transition temperatures for five different superconducting pairing symmetries for their corresponding coupling constants. The plots are shown in figure 1.

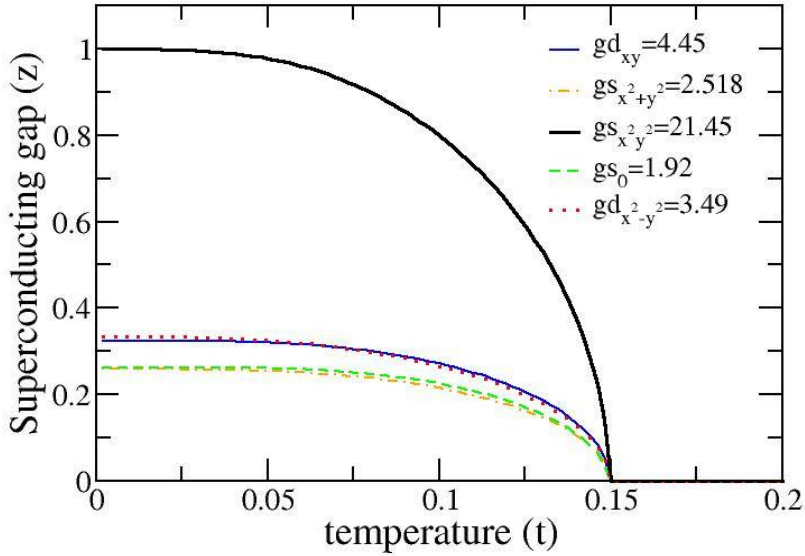

**Figure-1: The plot of superconducting gap (z) vs. temperature (t) for different waves of different values of g such as $gd_{xy} = 4.45$, $gs_{x^2+y^2}^2 = 2.518$, $gs_{x^2y^2}^2 = 21.45$, $gs_0 = 1.92$, $gd_{x^2-y^2}^2 = 3.49$, t1 = 1, t2 = -2.25, μ = 0.**

The SC transition temperature, $T_c$ nearly 30 K is observed for the superconductivity of the 122 systems [8]. For a given SC transition temperature $T_c$ = 0.15 ($T_c \approx 30$ K) corresponding to hopping parameter $t_1 = 0.02$ eV, the SC gap (z) is plotted with reduced temperature for SC coupling $gs_0 = 1.92$ for s-wave pairing, $gd_{x^2-y^2}$ = 3.49 for d-wave pairing and $gs_{x^2y^2} = 21.45$ for $s_\pm$ -wave pairing (see Fig 1). It is observed that the magnitude of the SC gap (z) is the smallest for s-wave pairing. The SC gap z value is slightly higher for d-wave pairing and the SC gap is the highest for $s_\pm$ wave pairing. All the temperature dependent SC gaps exhibit perfect mean-field behavior with $s_\pm$ symmetry showing the robust SC transition at $t_c \approx 0.15$. The relation $2\Delta_0(T) (T = 0 K)/k_B T_c \approx 3.52$ for s-wave pairing, $\approx 4.5$ for d-wave pairing and $\approx 13.3$ for $s_\pm$ -wave pairing, as compared to the BCS universal constant of 3.52 for isotropic metallic superconductors. It shows that this constant is much higher for $s_\pm$ pairing compared to BCS constant. This type of higher SC gap and $s_\pm$ symmetry are reported by ARPES experiments and the Local density approximation (LDA) calculations and band structure calculations [9, 10, 17, 31] and experiments. We have also computed the SC gap (z) for other two pairing symmetries with the SC couplings $gd_{xy}$ and $gd_{x^2-y^2}$. It is observed that the SC gap for coupling $gd_{xy}$ compares with the value of d-wave pairing symmetry with coupling $gd_{x^2-y^2}$. Further, the SC gap for the coupling $gd_{x^2-y^2}$ compares nearly with that of s-wave pairing with coupling $gs_0$.

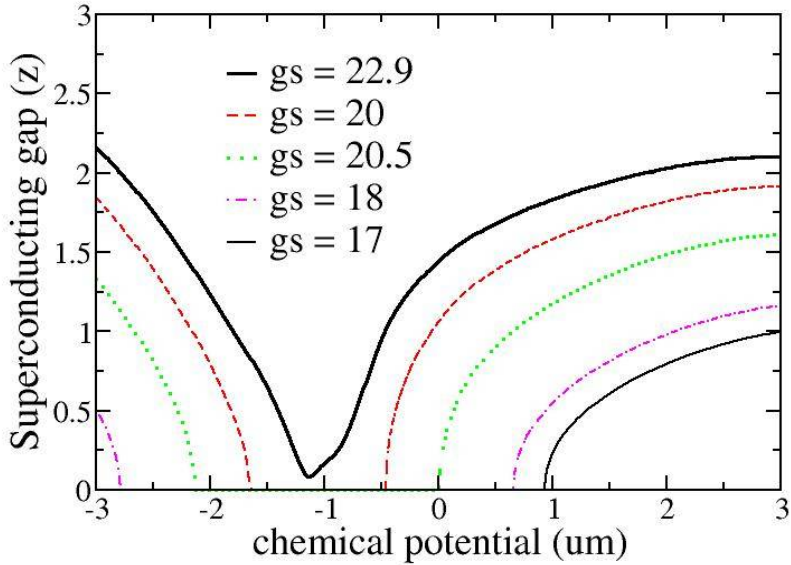

**Figure-2: The plot of superconducting gap (z) vs. chemical potential (μ) for different values of superconducting couplings gs = 22.9, 22, 20.5, 18, 17 for $s_\pm$ wave t1 = 1, t2 = -2.25, t = 0.1.**

Figure 2 shows the chemical potential dependence of SC gap for different values of SC coupling (gs) for $s_\pm$ wave pairing symmetry of the systems. For smaller SC coupling, gs = 17, we observe that SC gap increases with positive chemical potential with the minimum value of the chemical potential, um $\approx$ 0.9 below which there is no existence of superconductivity. On increasing the SC coupling, we observe that superconducting state exists for the two sets of values of chemical potential in the middle of which SC state doesn't exist. However, the SC state exists for all values of chemical potential for SC coupling gs = 22.9 and the SC gap becomes minimum at the chemical potential, um ≈ -1.2. There-fore we have considered a chemical potential of magnitude um ≈ 2.2 for our further calculations.

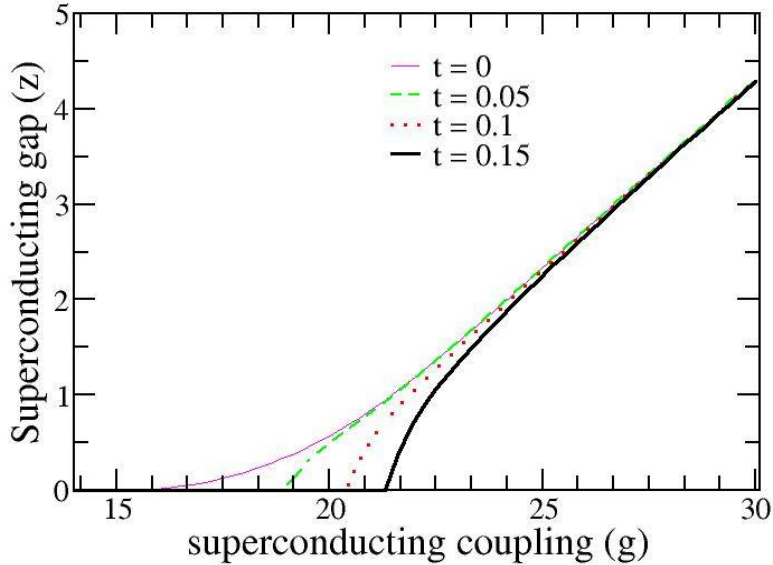

**Figure-3: The plot of superconducting gap (z) vs. superconducting coupling for different values of temperatures t = 0, 0.05, 0.1, 0.15 for $s_\pm$- wave, t1 = 1, t2 = -2.25, gs = 21.45.**

Figure 3 shows the plot of superconducting gap vs. superconducting coupling (g) for different values of temperatures t = 0, 0.05, 0.1, 0.15. We show how SC gap varies with the superconducting (SC) coupling (g) for different temperatures. It is observed that, for temperature t = 0, the minimum SC coupling becomes nearly 16.2 below which the SC phase vanishes. This minimum SC coupling increases with increase of temperature. The minimum SC coupling becomes nearly 21.45, for the temperature t = 0.15, ($T_c$ = 30 K) which is the SC transition temperature for $AFe_2As_2$ systems. There-fore we have considered the SC coupling $gs_{x^2y^2}$ ≈ 21.45 for $s_\pm$ superconducting pairing symmetry of the systems.

### 6.2 Study of electron specific heat

We have framed the free energy of the systems from which we have calculated numerically the temperature dependent entropy and electron specific heat for the systems. We have computed the electron specific heat for different values of SC couplings for $s_\pm$-wave pairing symmetry, which is very often observed in iron-based superconductors.

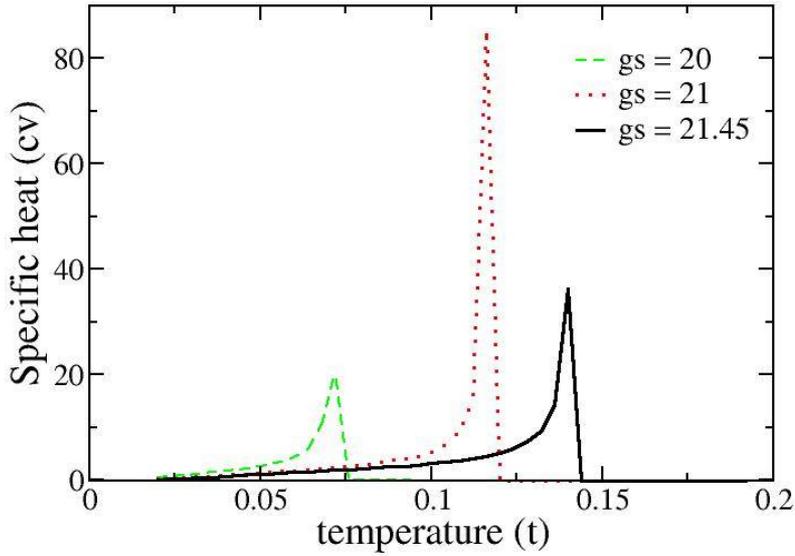

**Figure-4: The plot of specific heat ($c_v$) vs. temperature (t) for different values of superconducting couplings for $s_\pm$ wave, t1 = 1, t2 = -2.25, μ = 0.**

In figure 4, we show the temperature dependent electron specific heat for SC couplings gs = 20, 21 and 21.45 for $s_\pm$ pairing symmetry. For gs = 20, we observe that the electron specific heat nearly increases linearly with temperature at low temperatures and finally exhibits a sharp jump at SC transition temperature at $t_c$ = 0.075. When gs is increased, we observe similar temperature dependence of electron specific heat. However, the specific heat jump is shifted to higher temperatures with higher specific heat jumps at SC transition temperatures. The specific heat becomes very very small beyond SC transition temperature.

### 6.3 Study of electron density of states

The electron DOS for the system represents the tunneling conductance spectra of the system. Therefore, we have calculated electron DOS from the imaginary part of the Green's function and final expression is given in equation (25) and the DOS is computed numerically at a given temperature t = 0.1 for different pairing symmetries.

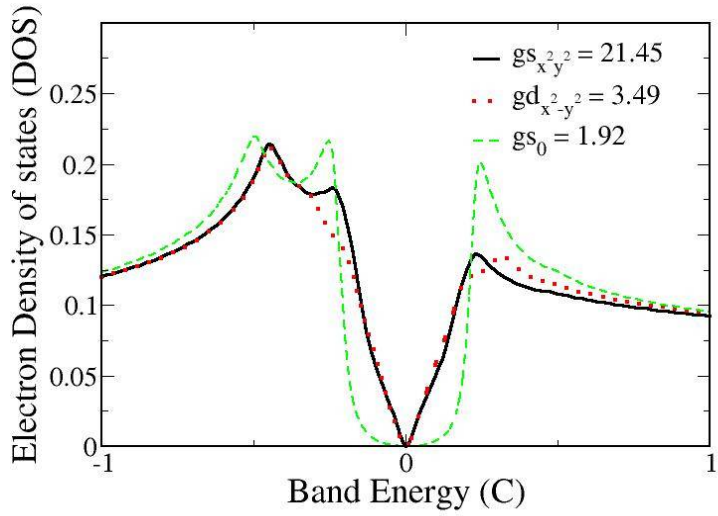

**Figure-5: The plot of density of states (DOS) vs. band energy (c) for all different waves without chemical potential μ, t1 = 1, t2 = -2.25.**

The different SC couplings are $gs_{x^2y^2} = 21.45$ for $s_\pm$-wave pairing symmetry, $gd_{x^2-y^2} = 3.49$ for d-wave pairing and $gs_0 = 1.92$ for $s_0$-wave pairing symmetry. Taking the SC gap parameter at temperature t = 0.1 the DOS is plotted as shown in figure 5. The figure 5 shows V-shaped SC gap for $s_\pm$-wave and d-wave pairing symmetries with asymmetric DOS, while the $s_0$-wave pairing shows U-shaped gap near the Fermi surface. The double peak structure appearing in the low energy edge of the asymmetric SC gap arises probably due to the next-nearest-electron hopping of the electrons in the 122-iron based systems.

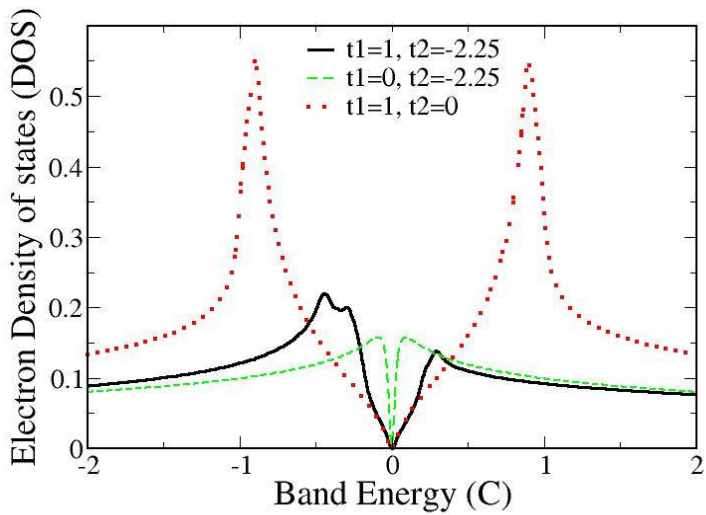

**Figure-6: The plot of the DOS vs. band energy (c) for varying values of t1 and t2 such as t1 = 1, t2 = -2.25; t1 = 0, t2 = -2.25; t1 = 1, t2 = 0; t = 0.1, μ = 2.2 for $s_\pm$-wave.**

In Fig 6, we study here the effect of the electron hopping parameters on the DOS. For t1 = 1 and t2 = 0 (i.e, only contribution of the nearest neighbor hopping), the DOS exhibits a V-shaped a symmetric SC gap with a node at the Fermi point (c = 0) and two sharp peak on either side of the Fermi surface (shown in dotted lines). For the contribution of the only second-nearest-neighbor-hopping (i.e, t2 = -2.25 and t1 = 0), the DOS is symmetrically V-shaped with a narrow gap and a node at Fermi point (shown in dashed lines). Further, the DOS is suppressed in this case. It appears that the major contribution to DOS is mainly due to the nearest-neighbor-hopping. The total contribution to DOS is shown in continuous line in figure 6. This shows an asymmetric V-shaped SC gap in DOS with a node. It is further observed that the peak in the DOS lying below the Fermi energy (i.e, at c = 0), splits into two due to the NNN-hopping of electrons to that of NN-hopping. This type of asymmetric gap structure is observed in the 122-type systems. However, Parish et.al. [32] have reported a U-shaped SC gap in their model calculation for $s_{\pm}$ pairing symmetry for their two band model calculations.

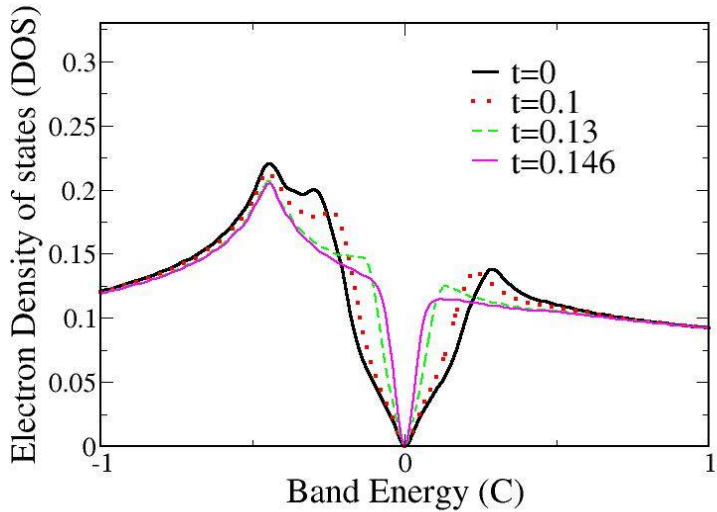

**Figure-7: The plot of the DOS vs. band energy (c) for $s_{\pm}$ wave for given temperatures t = 0, 0.1, 0.13, 0.146, t1 = 1, t2 = -2.25, μ = 0, gs = 21.45.**

The figure 7 shows the effect of temperature on the DOS i.e, on the conductance spectra of Fe-based superconductors taking the magnitudes of the SC gap for different pairings. As shown in figure 1, we find that the SC gap (z) gradually increases from SC transition temperature, $t_c$ = 0.15 to the temperature t = 0. Therefore, it is expected that the DOS should display the enhancement of the SC gap with decrease of temperature. For temperature t = 0.146 near $t_c$, the DOS shows a narrow SC gap with a node at Fermi point. The SC gap gradually widens with the decrease of temperature with the gap edges of the SC gap shifting away from the Fermi point. However, the DOS nearly remains constant far away from the Fermi point having constant electron DOS. The SC gap asymmetry is observed at all temperatures upto the SC transition temperature. However, the splitting of the low energy SC gap edge peak is prominent at lower temperatures.

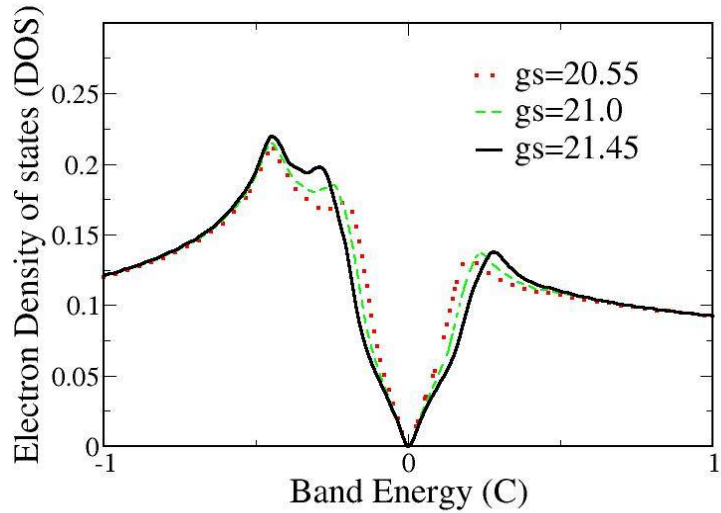

**Figure-8:** The plot of the DOS vs. band energy (c) for wave for different SC couplings, such as gs = 20.55, 21.0, 21.45, t1 = 1, t2 = -2.25, μ = 0.

In Fig 8, we study the effect of the $s_\pm$ pairing coupling on the electron DOS in absence of the chemical potential, since $s_\pm$ pairing mechanism is the most plausible mechanism observed in the iron based superconductors. The plots display the DOS of the system for different values of the couplings for $s_\pm$ pairing taking gs = 20.55, 21.0, 21.45. The DOS displays an asymmetric V-shaped gap structure in the DOS with a node at the Fermi level. Then the SC gap edge below the Fermi level exhibits the splitting of the peak with higher electron density of states compared to that above the Fermi level. With increase of the SC coupling, the V-shaped SC gap widens, while it is suppressed outside of the gap.

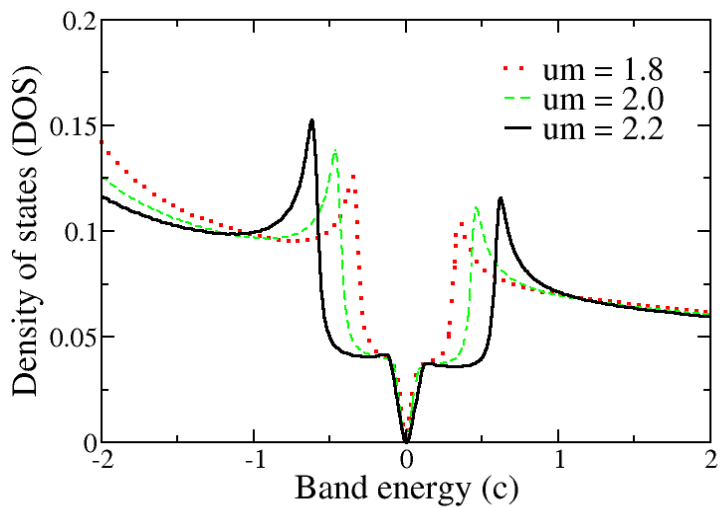

**Figure-9:** The plot of the DOS vs. band energy (c) for $s_\pm$ wave of different values of chemical potential μ = 1.8, 2.0, 2.2, t1 = 1, t2 = -2.25, t = 0.1, gs = 21.45.

In Fig. 9, we attempt to describe the effect of the chemical potential on the electron DOS taking different values of the reduced chemical potentials, um = 1.8, 2.0, 2.2. It exhibits an asymmetric two gap structure. The first SC gap in the DOS centered around the Fermi level, exhibits a narrow V-shaped gap with a node at the Fermi level. The second SC gap in the DOS, is an asymmetric wide U-shaped gap with finite electron density of states. The two gap structure probably arises due to the NN and NNN electron hoppings in the iron-based superconductors. It is observed that the SC gap in the DOS widens with increase of the chemical potential indicating that the magnitude of the SC parameter is enhanced. The enhancement of the SC gap arises due to the fact that more number of electrons participate in the superconducting pairing in the 122-type superconductors due to the increase of chemical potential (arising due to enhanced electron doping).

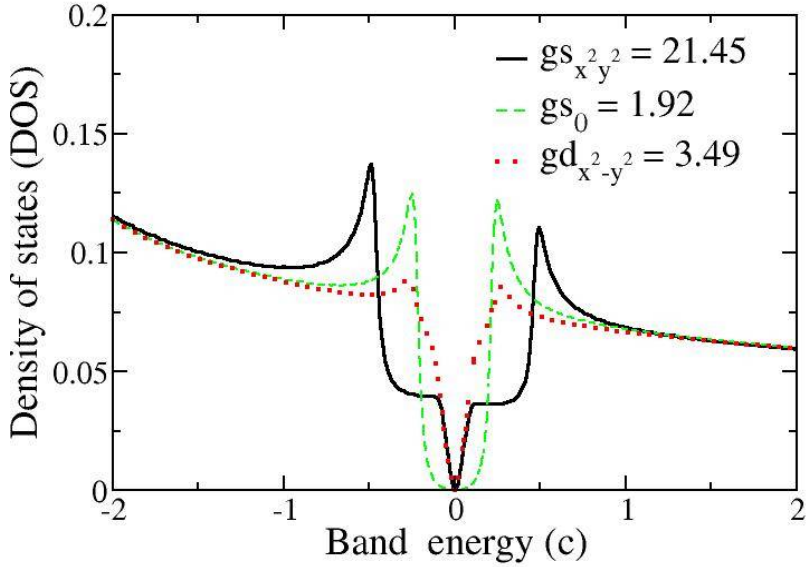

**Figure-10: The plot of the DOS vs. band energy (c) for all different waves for the value of chemical potential μ= 2.2, t1 = 1, t2 = -2.25, t = 0.1.**

Figure 10 shows the electron density of states (DOS) for the systems for different pairing symmetries in presence of chemical potential. For s-wave pairing with coupling $gs_0 = 1.92$, we observe a U –shaped SC gap and a V-shaped SC gap with a node at the Fermi point for the SC coupling $gd_{x^2-y^2} = 3.49$ for the d-wave pairing. On the other hand, we observe a completely different electron DOS for $s_\pm$ pairing symmetry with SC coupling $gs_{x^2y^2} = 21.45$. The DOS shows a perfect V-shaped SC gap with a node at the Fermi point, while it shows another asymmetric wide U-shaped SC gap. With higher gap width, it is appearing as if the DOS structure is a combined effect of the d-wave as well as the s-wave pairing symmetries. This type of peculiar gap structure $s_\pm$ pairing for the 122-type iron-based systems arises perhaps due to the NN and NNN- hopping of electrons in the doped condition with a finite chemical potential. This conclusion will be more obvious latter from the discussions in Fig 11. This type of anomalous gap structure is observed in the $AFe_2As_2$ type iron based superconductors, as reported earlier by K. Seo et.al. [31].

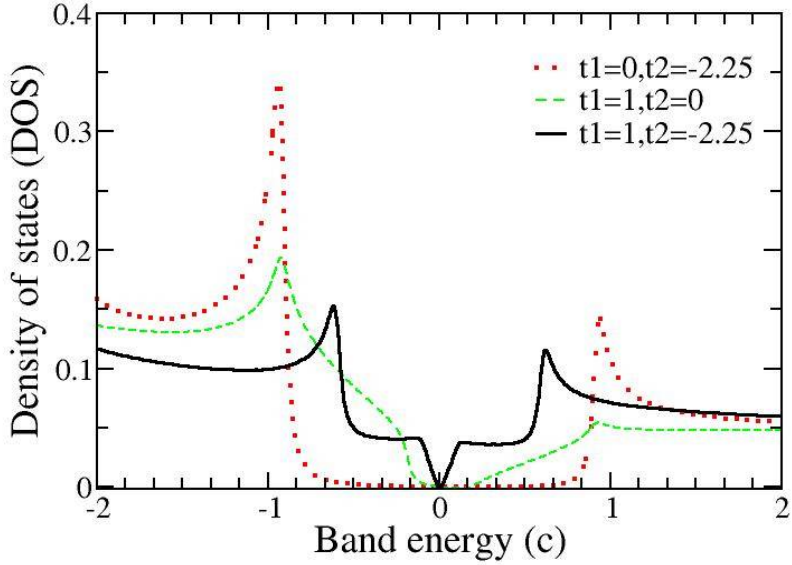

**Figure-11: The plot of the DOS vs. band energy (c) for $s_\pm$ wave for other physical parameters like t1 and t2 all are in one, gs = 21.45, μ= 2.2.**

Figure 11 shows the effect of the electron hopping integrals on the electron DOS for $s_\pm$-wave pairing in presence of chemical potential. For second-nearest-neighbor electron hopping integral t2 = -2.25, in absence of nearest-neighbor electron hopping (t1 = 0), the DOS shows the the U-shaped gap with a very strong asymmetry in it (shown in dotted lines). For nearest-neighbor electron hopping t1 = 1, in absence of second-nearest-neighbor electron hopping integral (t2 = 0), the DOS exhibits an asymmetric gap and the electron density below the Fermi surface becomes large as compared to that above it (shown by dashed lines). Finally the electron density is plotted in presence of a finite of t1 and t2. It shows two gap structures with reduced asymmetric gap structure. However, the DOS exhibits a small V-shaped gap near the Fermi surface with a node. Thus it is obvious the gap structure in the DOS of the 122-type iron based systems for the $s_\pm$ pairing symmetry arises due to the combined effect of the two types of electron hoppings in presence of a finite chemical potential. This type of gap structure has been reported for 122-type systems by K. Seo et.al. [31] and Parish et.al. [32].

## 7  Conclusions

We have attempted here to investigate the mechanism of the superconducting (SC) pairing symmetry in the 122-type iron based superconductors. The Hamiltonian consists of the nearest-neighbor (NN) as well as the next-nearest-neighbor (NNN) electron hopping interactions as well as the superconducting (SC) interaction for the 122 systems. Here the SC gap parameter is electron momentum dependent associated with different possible pairing symmetries. The SC gap equation is calculated by Zubarev's Green's function technique and is solved self-consistently. Further, the tunneling spectra is predicted from the calculated electron density of states (DOS). The temperature dependent SC gap shows that the magnitude of the SC gap is the largest for the $s_\pm$ pairing symmetry in 122-type superconductors as compared to that of the d-wave pairing and others. For the doped 122 systems, the SC gap becomes finite for all values of chemical potentials representing both electron and hole-dopings for values of the SC coupling associated with the $s_\pm$ pairing symmetry. It is observed that the SC gap vanishes for a given temperature below a critical value of the SC coupling and this critical coupling increases from 0 K to the SC transition temperature. The density of state in absence of the chemical potential representing the tunneling conductance spectra exhibits an asymmetric V-shaped SC gap for the $s_\pm$ pairing symmetry with a node at the Fermi level and a splitting of the peak lying at the lower energies. This peak splitting indicates the occurrence of the two SC gap structure arising due to NN and NNN hoppings in the 122 systems. This type of gap structure has been reported by Seo et. al. [31] for the 122-type iron based superconductors like $MFe_2As_2$ with M representing an alkali ion. The doping of the system with finite chemical potential displays two gap structure in the DOS for $s_\pm$ pairing and exhibits a narrow V-shaped gap with a node as well as an asymmetric, U-shaped gap with finite electron density similar to that observed by Seo et. al. [31] and by Parish et. al [32] for the two band model calculation for the systems. This present work is published [33, 34].